\documentclass{aa}
\usepackage{graphicx,natbib,amsmath,afterpage}

\newcommand{\an}{{Astr.~Nachr.}}
\newcommand{\Sci}{{Science}}
\newcommand{\ASS}{{Astrophys.\&~Space~Sci.}}

\newcommand{\Hskin}{H_{\mbox{\scriptsize skin}}}
\newcommand{\omegacyc}{\omega_{\mbox{\scriptsize cyc}}}
\newcommand{\rbcz}{r_{\mbox{\scriptsize bcz}}}
\newcommand{\rin}{r_{\mbox{\scriptsize in}}}

\newcommand{\rt}{r_{\mbox{\scriptsize t}}}
\newcommand{\Omegabcz}{\Omega_{{\mbox{\scriptsize bcz}}}}
\newcommand{\Prm}{\mbox{Pr}_{{\mbox{\scriptsize m}}}}
\newcommand{\Pcyc}{P_{\mbox{\scriptsize cyc}}}

\addtocounter{totalnumber}{4}
\addtocounter{topnumber}{4}

\defcitealias{FD+P:AA02}{Paper~I}

\begin{document}
   \title{Dynamics of the fast solar tachocline}

   \subtitle{II. Migrating field}

   \author{E. Forg\'acs-Dajka\inst{1,2}}

   \offprints{E. Forg\'acs-Dajka}

   \institute{\inst{1}E\"otv\"os University, Department~of Astronomy, Budapest, Pf.~32,
              H-1518 Hungary\\
              \inst{2}Konkoly Observatory, Budapest, Pf.~67,
              H-1525 Hungary\\
              \email{E.Forgacs-Dajka@astro.elte.hu}
             }

   \date{Astronomy and Astrophysics, v.413, p.1143-1151 (2004)}

\abstract{

We present detailed numerical calculations of the fast solar tachocline based on 
the assumption that  the dynamo field dominates over 
the dynamics of the tachocline. In the present paper of the series, 
we focus on three shortfalls of the earlier models. First, instead of 
the simple oscillating dipole poloidal field we study the more general magnetic 
field structures reminiscent of the butterfly diagram. 
The migrating field is prescribed as  the observed axisymmetric radial magnetic field 
\citep{Stenflo:ASS88,Stenflo:SSM94}. Our results are in good agreement with our analitical estimate 
and our previous works in \citet{FD+P:SolPhys01,FD+P:AA02}, but the polar ''dip'' in isorotational surfaces is 
strongly reduced in this case. 
On the other hand, a more realistic model should have a magnetic diffusivity decreasing significantly inside 
the radiative interior, so we also explore the effect of diffusivity and magnetic Prandtl number 
varying with depth. We found that the downwards decreasing magnetic diffusivity and Prandtl number 
have no significant effect on the solution, although the temporal variation of the tachocline thickness 
has decreased. 

}

   \maketitle
%

\section{Introduction}

Helioseismic data allow us to study the structure and rotation
rate of the solar interior. The tachocline is a thin
transitional layer below the solar convective zone. In this layer
the surface-like rotation of the convection zone changes to
near-uniform rotation in the radiative interior.  The existence
and properties of this layer have been known from helioseismic
studies, recently reviewed by \citet{Corbard+:SOGO}. The
tachocline is known to be extremely thin; the central radius of
the tachocline appears to lie within the range
$0.69-0.71\,R_\odot$, and its thickness is estimated to be 
$<0.05\,R_\odot$ \citep{Kosovichev:ApJ96,Corbard+:AA99,
Schou+:ApJ98,Charbonneau+:ApJ99}.

Precise values for the parameters of the solar tachocline as such
mean position, thickness, latitudinal and temporal
variations, depend on the inversion techniques used for
helioseismic data. \citet{Elliott+Gough:ApJ99} suggest that there 
exists a difference in the sound speed profile between the observations and the 
Standard Solar Models. By including an additional mixing layer below 
the convection zone, they calibrate the thickness of the tachocline and 
they find that the mean tachocline thickness is $0.019\,R_\odot$ with a formal 
standard error of about $5$\%. The observed light-element 
abundances at the surface also suggest that the mixing process in the solar tachocline 
is confined to a shallow layer \citep{Brun+:ApJ99,Brun+:AA02}.

Making use of the most recent helioseismic results one can investigate the temporal variation 
and the latitudinal structure of the tachocline. \citet{Basu+Antia:MNRAS01} found that 
the tachocline is prolate, with the difference between the tachocline position at 
$0^\circ$ and that at $60^\circ$ latitude being about $0.02\,R_\odot$. This is in 
agreement with results obtained by \citet{Charbonneau+:ApJ99}. \citet{Basu+Antia:MNRAS01} 
also report an increase in thickness of the tachocline with latitude, but this 
increase is less significant, though still at the $3\sigma$ level. Regarding the 
dynamical aspect, one would expect temporal variations associated with the solar cycle 
to appear in the tachocline \citep{Howe+:Sci00,Basu+Antia:MNRAS01,Corbard+:SOGO}.
\citet{Howe+:Sci00} report that a large-scale oscillation may be 
taking place in this layer with a period about $1.3$ yr. It is not clear 
whether this period is associated with solar-cycle-related variations, and 
\citet{Basu+Antia:MNRAS01} did not find any periodic or systematic changes in 
rotation rate in the tachocline region.  
This fact may be the consequence of the insufficiency of the helioseismic
data.

The extreme thinness of the tachocline implies a strongly anisotropic
transport of angular momentum \citep{Spiegel+Zahn:AA92}. Several different
mechanisms have been proposed for this transport, but it is now widely believed
that the magnetic field is instrumental in its origin. 
It is known that an oscillating magnetic field cannot propagate far into 
the radiative zone. The extent of the penetration of this magnetic field is the skin depth 
$\Hskin=\sqrt{2\eta/\omegacyc}$, where $\eta$ is the magnetic diffusivity and 
$\omegacyc$ is the frequency of the cycle. On the one hand, for $\eta\la 10^8\,$cm$^2/$s 
the dynamo field cannot penetrate the tachocline, and we can expect the 
tachocline to be pervaded by the internal primordial field. On the other hand, 
for $\eta\ga 10^9\,$cm$^2/$s the tachocline dynamics should be governed by the dynamo field.
As the associated diffusive and Lorentz timescales are also very different, 
these two cases basically correspond to the case of ``slow'' and ``fast'' tachocline, 
discussed in the literature \citep[see esp. Table I in][]{Gilman:SolPh00}.

In recent years the case of a slow tachocline has been investigated extensively by 
a number of authors.  
\citet{Rudiger+Kichatinov:AN97} and \citet{McGregor+Charbonneau:ApJ99} studied the 
interaction between a large-scale field with fixed poloidal component and differential rotation 
without taking into consideration the meridional circulation. They found that 
the internal poloidal field of $10^{-3}\,$G is sufficient to confine the tachocline to its
observed thickness if the internal field is fully contained within the
radiative zone.
\citet{Gough+McIntyre:Nat98} presented a model for the solar 
tachocline which allows the nonlinear interaction between the 
meridional flows and a large-scale magnetic field in self-consistent way. 
\citet{Garaud:MNRAS01} performed 
calculations taking into account the meridional flow and the self-consistent evolution of
the poloidal field. 

In the first paper of these series \citep[][hereafter Paper I]{FD+P:AA02}, 
we studied the alternative case of a fast, 
turbulent tachocline with a turbulent diffusivity
of about $\eta=10^{10}\,$cm$^2$/s and we presented detailed 
numerical calculations allowing for the self-consistent evolution of
the dipolar poloidal field. It was found that a sufficiently 
strong oscillatory poloidal field with dipolar latitude dependence at 
the tachocline-convective zone boundary is able to confine the tachocline to its observed
thickness. This is in good agreement with our analitical estimate in \citet{FD+P:SolPhys01}.

In this paper, instead of a dipolar field we consider the more realistic case of 
a migrating field, studying its effect on the dynamics of the solar tachocline. 
After formulating the mathematical problem in Sect.\,2, results of our model 
containing a migrating magnetic field
will be presented in Sect.\,3
We also explore the effect of the meridional 
flow and the impact of varying the diffusivity with radius.
Finally, Sect.\,4 summarizes the main results.


\section{The model}

In order to describe the differential rotation and the 
evolution of the large-scale magnetic field in 
the solar interior it is useful to write the equation of motion  and 
the hydromagnetic induction equation in 
a frame rotating with the fixed internal rotation rate. These 
equations can be found in the first paper of these series 
(\citetalias[][Sect.~2.1]{FD+P:AA02}).

\subsection{Flow Field}

The flow field was treated in the same manner as in \citetalias{FD+P:AA02}, nevertheless,  
some additional remarks are taken here.
As we mentioned in \citetalias[][Sect.~3.4]{FD+P:AA02}, the total flow field can be written as  
the sum of the differential rotation and meridional circulation. 
While we compute the evolution of the angular velocity with time, 
the meridional circulation is prescribed in our region, because 
in this case we focus on the effect of any meridional circulation in the
tachocline, not on the complex problem of meridional circulation 
in the convective zone. Recent 
helioseismic inversions have shown that the outer convective 
envelope is pervaded by a $\simeq 20 \mbox{ m s}^{-1}$ poleward 
meridional flow. Such inversions currently do not provide accurate 
meridional flow information at greater depth, thus we turn for 
guidance to the numerical simulations of thermally driven turbulent 
convection recently carried out by \citet{Miesch+:ApJ00}, 
\citet{Kuker+Stix:AA01} and \citet{Kitchatinov+Rudiger:AA95}. 
These simulations generate a mean meridional 
circulation and in these the meridional flow penetrates into the stable interior. 
In an attempt to incorporate at least qualitatively these observational 
and computational results, we used a model similar to that described in 
\citetalias{FD+P:AA02}. 

The strong subadiabatic stratification of the solar interior sets 
an upper limit to the meridional flow. An elementary estimate discussed 
in \citetalias{FD+P:AA02} yields $v_m \sim 10 \mbox{ cm s}^{-1}$. 
On the other hand \citet{Miesch+:ApJ00} found that depending on assumed values of some of the input 
parameters this pattern is a persistent equatorward or poleward 
circulation in the overshoot region. 
Accordingly, the flow parameters used in our calculations were chosen to produce simple
smooth one- and two-celled flow patterns obeying the amplitude constraint
as before mentioned, while reproducing the observed flow speed near the surface. 
For the density, we used the following expression:
\begin{eqnarray}
\rho(r)=C\left(\frac{R_\odot}{r}-\gamma\right)^m,
\end{eqnarray}
where the values of the parameters used are $C=1.2$, $m=2.0$ and $\gamma=0.9$.
The flow patterns are shown in \citetalias[][Figs.~9 and 10]{FD+P:AA02}. In these cases, the speed of the horizontal motion is $\sim
5\,$cm/s in the upper part of the radiative interior. 

\subsection{Poloidal Field}

   \begin{figure}
   \centering
   \includegraphics[width=1.03\linewidth]{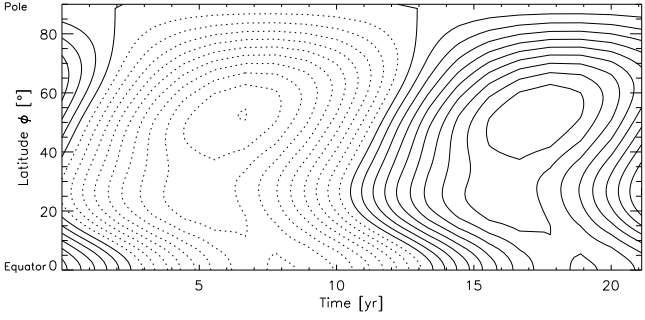}
      \caption{Illustration of the evolution of the vector potential. 
               Solid lines represent positive $A$ and dashed lines denote 
	       negative $A$.
              }
         \label{fig:A_time}
   \end{figure}

It is well known that sunspot activity occurs in the form of waves with periods 
$\sim11$ yr, which propagate from middle latitudes towards the solar equator. In the extended 
butterfly diagram \citep{Makarov+Sivaraman:SolPhys89}, one can see the poleward migration of 
the magnetic fields indicated by the migration pattern of a number of tracers such as 
quiescent prominences or the coronal green line. 

\citet{Stenflo:ASS88,Stenflo:SSM94} analyzed  
$33$ yr of synoptic observations of the Sun's magnetic field carried out daily at the Mount Wilson 
and Kitt Peak observatories over all solar latitudes and longitudes. The observed longitudinal 
magnetic field has been converted to a radial magnetic field assuming that the field 
direction is on average radial in the layers where the field is measured. 
This provides an opportunity to analyse the global modes of the Sun.
\citet{Stenflo:ASS88,Stenflo:SSM94} expanded the radial 
magnetic field in spherical harmonics and he found that the zonal magnetic field pattern can be  
represented as a superposition of $N$ discrete modes with purely sinusoidal time variations 
with frequency $\omegacyc=2\pi/22$ yr$^{-1}$. Consequently, the evolutionary pattern of 
the axisymmetric radial magnetic field can be written 
in the following form:
\begin{eqnarray}
B_r(t,x)=\sum_{l=1}^N |a_l|\cos\left[\omegacyc\left(t-t_l\right)\right]P_l(x),
\end{eqnarray}
with
\begin{eqnarray}
\Phi_l=-\omegacyc t_l,
\end{eqnarray}
where $B_r$ is the radial, axisymmetric magnetic field, $|a_l|$ is the amplitude, 
$\Phi_l$ is the phase lag, $t_l$ is the time lag and $x=\cos\theta$.

\cite{Petrovay+Szakaly:1999} found that the 
latitudinal distribution of the field at the surface reflects the conditions at 
the bottom of the convective zone, i.e.\, in this regard the convective 
zone behaves as a ``steamy window''.
Thus, if we assume that the above formula describes the time-evolution of the radial magnetic field 
at the base of the convective zone, then the development of the vector potential is given by: 
\begin{eqnarray}
\label{eq:polter}
A(r,\theta,t)=\frac{\int B_r(r,\theta,t)\sin\theta r \mathrm{d}\theta + F(r,t)}{\sin\theta},
\end{eqnarray}
where $F$ is an arbitrary function.  
\citet{Stenflo:ASS88,Stenflo:SSM94} also studied the dominance of the odd and even modes and  
he showed that the odd modes prevail in the evolution of the radial, axisymmetric magnetic field. 

Thus, the development of the vector potential is represented as a superposition of the sinusoidal, 
$22$ yr variations for the $7$ odd modes with $l=1,3,\dots,13$, as
\begin{eqnarray}
\begin{aligned}
A&(\theta,t)=A_0\sum_{k=0}^6 V_{1+2k}(\theta)|a_{1+2k}|\cos\left[\omegacyc\left(t-t_{1+2k}\right)\right] \\
&V_1=\frac{1}{2}\sin\theta \\
&V_3=\frac{1}{2}\sin\theta-\frac{5}{8}\sin^3\theta \\
&V_5=\frac{1}{2}\sin\theta-\frac{7}{4}\sin^3\theta+\frac{21}{16}\sin^5\theta \\
&V_7=\frac{1}{2}\sin\theta-\frac{27}{8}\sin^3\theta+\frac{99}{16}\sin^5\theta-\frac{429}{128}\sin^7\theta \\
&V_9=\frac{1}{2}\sin\theta-\frac{11}{2}\sin^3\theta+\frac{143}{8}\sin^5\theta-\frac{715}{32}\sin^7\theta \\
    &\quad +\frac{2431}{256}\sin^9\theta \\
&V_{11}=\frac{1}{2}\sin\theta-\frac{65}{8}\sin^3\theta+\frac{325}{8}\sin^5\theta-\frac{5525}{64}\sin^7\theta \\
    &\quad +\frac{20995}{256}\sin^9\theta-\frac{29393}{1024}\sin^{11}\theta \\
&V_{13}=\frac{1}{2}\sin\theta-\frac{45}{4}\sin^3\theta+\frac{1275}{16}\sin^5\theta-\frac{8075}{32}\sin^7\theta \\
    &\quad +\frac{101745}{256}\sin^9\theta-\frac{156009}{512}\sin^{11}\theta+\frac{185725}{2048}\sin^{13}\theta.
\end{aligned}
\end{eqnarray}
The factor $A_0$ fixes the poloidal field amplitude.
The $|a_{l}|$ amplitudes and $\Phi_l$ phases of the 
odd modes were obtained by \citet{Stenflo:SSM94}. The result is shown in Fig.\,\ref{fig:A_time}.

\subsection{Equations for Our Model}

   \begin{figure*}[!ht]
   \centering
   \includegraphics[width=0.60\linewidth]{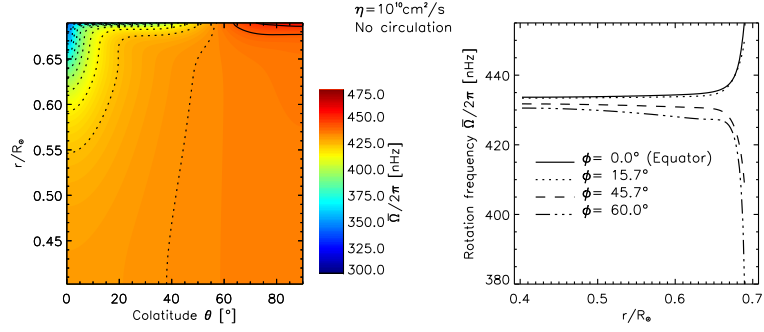}
      \caption{Spreading of the differential rotation into the radiative
	       interior for $\eta=\nu=10^{10} \mbox{cm}^2/\mbox{s}$.  {\it
	       Left-hand panel:} contours of the time-average of the 
	       rotation rate under one dynamo period, with contour spacing 
	       $8.75\,\mbox{nHz}$; the solid lines represent the regions 
	       rotating faster and the dotted lines show the regions 
	       slower than $\Omega_0/2\pi=437$nHz. 
	       {\it Right-hand panel:} the rotation rate 
	       at different latitudes as a function of
	       radius.  The peak amplitude of the poloidal magnetic field and 
	       the toroidal field are $B_p\sim2400$ G and $B\sim12000$ G, respectively.}
         \label{fig:1_atlag_col}
   \end{figure*}

Based on \citetalias[][Sect.~2.1]{FD+P:AA02}, the azimuthal components
of Eqs.~(1--2), including the effects of
diffusion, Coriolis force, meridional circulation, toroidal field production by
shear, and Lorentz force, read 
\begin{eqnarray}
\label{eq:main1}
\partial_t \omega &=& \partial_r \omega \left( \partial_r \nu + 4 \frac{\nu}{r} 
                                       + \partial_r \rho \frac{\nu}{\rho} \right) +
                      \partial_\theta \omega \frac{3\nu\cos\theta}{r^2\sin\theta} + \\
		   && \partial^2_r \omega \nu +
		      \partial^2_\theta \omega \frac{\nu}{r^2} +
		      \mbox{L} + \mbox{C} + \mbox{M$_1$}  \nonumber \\
\mbox{L} &=& \frac{B}{4\pi\rho r^3 \sin^2\theta}\left(\partial_\theta A \sin\theta - \partial_r A r \cos\theta \right) +  \nonumber \\
   && \frac{\partial_r B}{4\pi\rho r^2 \sin^2\theta}\left(\partial_\theta A \sin\theta - A \cos\theta \right) +  \nonumber \\
   && \frac{\partial_\theta B}{4\pi\rho r^3 \sin^2\theta}\left(-A \sin\theta - \partial_r A r \sin\theta \right) \nonumber \\
\mbox{C} &=& \psi \frac{2\Omega_0}{r^3 \rho}\, ( 1-3 \cos^2\theta) + 
      \partial_r \psi \frac{2\Omega_0\cos^2\theta }{r^2 \rho} \nonumber \\
\mbox{M$_1$}&=& \omega \left[ \psi \left(1-3\cos^2\theta\right) + \partial_r \psi r \cos^2\theta\right]\frac{2}{r^3\rho} + \nonumber \\
    && \partial_r\omega \psi \frac{1-3\cos^2\theta}{r^2\rho} + 
       \partial_\theta \omega \partial_r\psi \frac{\sin\theta\cos\theta}{r^2\rho} \nonumber
\end{eqnarray}
\begin{eqnarray}
\label{eq:main2}
\partial_t B &=& B \left( \frac{\partial_r \eta}{r} -\eta \frac{\cos^2\theta}{r^2\sin^2\theta} - \frac{\eta}{r^2} \right) + \\
             &&  \partial_r B \left( \frac{2\eta}{r} + \partial_r \eta \right) +
		 \partial_\theta B \frac{\eta\cos\theta}{r^2\sin\theta} + \nonumber \\
	     &&  \partial^2_r B \eta + 
		 \partial^2_\theta B \frac{\eta}{r^2} +  \nonumber \\
	     &&  \partial_r \omega \left( \partial_\theta A \sin\theta + A \cos\theta \right) + \nonumber \\
             &&  \partial_\theta \omega \left( -A \sin\theta - \partial_r A r \sin\theta \right) \frac{1}{r^2} +
	         \mbox{M$_2$} \nonumber \\
\mbox{M$_2$} &=& B \left[ \psi\left(3\cos^2\theta-1\right)\left(\frac{1}{r}+\frac{\partial_r \rho}{\rho}\right)- \right. \nonumber \\
              &&          \left. \partial_r\psi\cos^2\theta\right]\frac{1}{r^2\rho} +  \nonumber \\
              && \partial_r B \left(1-3\cos^2\theta\right)\frac{\psi}{r^2\rho} + \nonumber \\
	      && \partial_\theta B \frac{\sin\theta\cos\theta}{r^2\rho}\partial_r \psi \nonumber
\end{eqnarray}
\begin{eqnarray}
\label{eq:main3}
\partial_t A &=& A \frac{-\eta}{r^2\sin^2\theta} + 
                 \partial_r A \frac{2\eta}{r} + 
		 \partial_\theta A \frac{\eta\cos\theta}{r^2\sin\theta} + \\
	     &&  \partial^2_r A \eta + 
		 \partial^2_\theta A \frac{\eta}{r^2} +   
		 \mbox{M$_3$} \nonumber \\
\mbox{M$_3$} &=& A \left[ \psi\left(1-3\cos^2\theta\right) + \partial_r\psi r \cos^2\theta \right]\frac{1}{r^3\rho} + \nonumber \\
              && \partial_r A \psi \left(1-3\cos^2\theta\right) \frac{1}{r^2\rho} + 
		 \partial_\theta A \frac{\sin\theta\cos\theta}{r^2\rho}\partial_r \psi \nonumber 
\end{eqnarray}
(For the sake of completeness we repeated here Eqs.\,($13-15$) of \citetalias{FD+P:AA02} which contain a typo.)
In these equations, $L$ represents the terms associated with the Lorentz force, $C$ denotes the
terms  associated with the Coriolis force and M$_1$, M$_2$, M$_3$ denote the
terms associated  with the advection by meridional circulation.

   \begin{figure}[!h]
   \centering
   \includegraphics[width=0.7\linewidth]{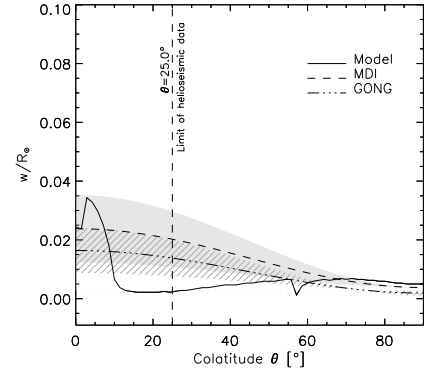}
      \caption{Latitudinal variation of the tachocline thickness for the
	       case in  Fig.\,\ref{fig:1_atlag_col}. The solid line represents 
	       the thickness from the model and the dashed and dashed-dot lines 
	       show $w$ from MDI and GONG data respectively \citep{Basu+Antia:MNRAS01}.
               The grey and the hatched areas show the error estimates on these.
	      }
         \label{fig:1_w_th}
   \end{figure}

   \begin{figure}[!h]
   \centering
   \includegraphics[width=0.7\linewidth]{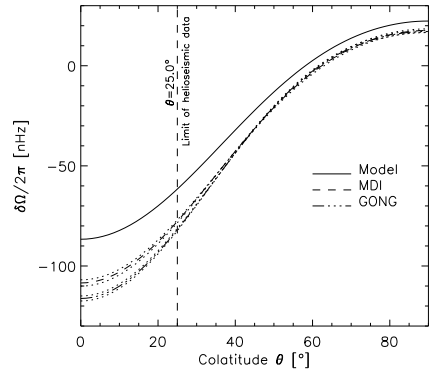}
      \caption{Latitudinal variation of the radial jump in $\Omega$ for the
	       case in  Fig.\,\ref{fig:1_atlag_col}. The solid line represents 
	       the thickness from the model and the dashed and dashed-dot lines 
	       show $\delta\Omega$ from MDI and GONG data respectively \citep{Basu+Antia:MNRAS01}.
               The dotted lines show the error estimates on these.
              }
         \label{fig:1_jump_th}
   \end{figure}

\subsection{Boundary Conditions for the Physical Parameters}

Our model does not include the convection zone, so 
the computational domain for the present calculations consists of just an annular 
meridional cut in the northern hemisphere $(0\le\theta\le\pi/2)$ in 
the upper part of the radiative interior, between radii $\rin$ and $\rbcz$, where
$\rbcz$  is the radius of the bottom of the convection zone. For the
solution of Eqs.\,(\ref{eq:main1}--\ref{eq:main3})
we use the following boundary conditions.

Assuming the deep radiative core to be a rigidly rotating perfect conductor, we 
set both $B$ and $A$ to zero at the lower boundary. Both these quantities are also 
set to zero along the symmetry axis and they are antisymmetric about the equator. 
Furthermore, our upper boundary condition on the poloidal field is the solution 
of Eq.~(\ref{eq:polter}). For the upper boundary condition chosen for
the toroidal field we assume that the toroidal field is negligible in the
convective zone compared to its value in the tachocline. 
Such a hypothetical
situation is in line with current thinking in dynamo theory, and it may be a
natural consequence of buoyancy-driven instabilities effectively removing any
toroidal flux from the convective zone.

For $\omega$ we require axial and equatorial symmetry, and 
at the bottom of the convection zone we suppose that 
the rotation rate can be described with the same
expression as in the upper part of the convection zone. 
Since the observations 
indicate that the differential rotation varies little within the convection 
zone and in the deep radiative interior, we can take $\partial_r\omega=0$ at the bottom 
of the computational domain, so  
the lower boundary condition for $\omega$ is different than in 
\cite{FD+P:SolPhys01,FD+P:AA02} where $\omega=0$ at the bottom of our box. 
Finally, we summarize and formulate these conditions in this way:
\begin{eqnarray}
\label{eq:boundary}
\begin{array}{lll}
A &= A_0\sum_{k=0}^6 V_{1+2k}|a_{1+2k}|&\cos\left[\omegacyc\left(t-t_{1+2k}\right)\right] \\
				       &&\mbox{at } r = \rbcz \\ 
\omega &= \Omegabcz-\Omega_0   &\mbox{at } r = \rbcz \\
B &= 0   &\mbox{at } r = \rbcz \\
B &= A = \partial_r\omega = 0   &\mbox{at } r = \rin \\
B &= A = \partial_\theta\omega = 0  &\mbox{at } \theta = 0, \\ 
B &= \partial_\theta A = \partial_\theta\omega = 0   &\mbox{at } \theta = \pi/2, 
\end{array}
\end{eqnarray}
where $\Omega_0$ is the rigid rotation speed at the observed rate $\Omega_0/2\pi = 437$ nHz 
and $\Omegabcz$ is the rotation rate, which can be described with the same
expression as in the upper part of the convection zone. 

In accordance with the
observations of the GONG network, the expression used for
$\Omegabcz$ is
\begin{eqnarray}
\frac{\Omegabcz}{2\pi} = 456 - 72 \cos^2\theta - 42 \cos^4\theta 
  \hspace{0.2cm} \mbox{nHz}.
\end{eqnarray}

\subsection{Solution procedure}

The Navier-Stokes, the induction equations and the equation for 
the poloidal magnetic field are coupled, nonseparable partial differential 
equations. We solve them as an initial-boundary value problem. 
We use a time relaxation method with a finite difference scheme first order
accurate in time to solve the equations. A uniformly spaced grid is set up
with 128 grid points in the $r$ direction and 64 grid points in the $\theta$ direction.

The initial conditions chosen for all calculations are 
\begin{eqnarray}
\label{eq:incond}
\begin{array}{lll}
A (r,\theta ,t=0) &= A_0\sum_{k=0}^6 V_{1+2k}|a_{1+2k}|&\cos\left(\Phi_{1+2k}\right) \\
			&&\mbox{at } r = \rbcz\\
A (r,\theta , t=0)     &= 0 & \mbox{at } r < \rbcz \\
\omega (r,\theta ,t=0) &= \Omegabcz - \Omega_0 & \mbox{at } r = \rbcz\\
\omega (r,\theta ,t=0) &= 0  & \mbox{at } r < \rbcz \\
B (r,\theta ,t=0)      &\equiv  0 & 
\end{array}
\end{eqnarray}
Starting from the initial conditions, 
the solution is allowed to evolve in time until it relaxes to a very nearly periodic
behaviour.

   \begin{figure}
   \centering
   \includegraphics[width=0.7\linewidth]{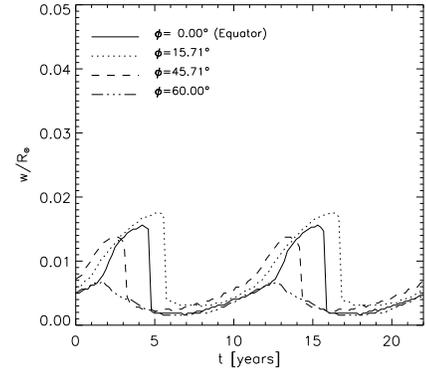}
      \caption{The thickness of the tachocline at different latitudes as a 
               function of time
	       for the case in  Fig.\,\ref{fig:1_atlag_col}.
	       }
         \label{fig:1_w_time}
   \end{figure}

   \begin{figure}[b]
   \centering
   \includegraphics[width=1.03\linewidth]{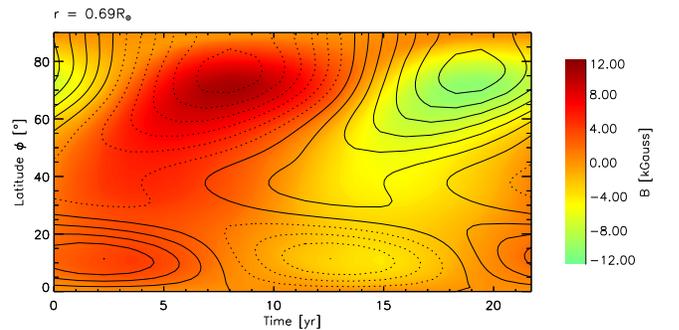}
      \caption{Time-latitude diagram for the toroidal field ({\it shaded region})  
               and the radial field ({\it contours}) at the base of the convection zone. 
	       Equidistant contour levels of the radial field are 
	       separated by intervals of $30\,$G.
      	      }
         \label{fig:1_tor_rad_time}
   \end{figure}

   \begin{figure*}[!ht]
   \centering
   \includegraphics[width=0.49\linewidth]{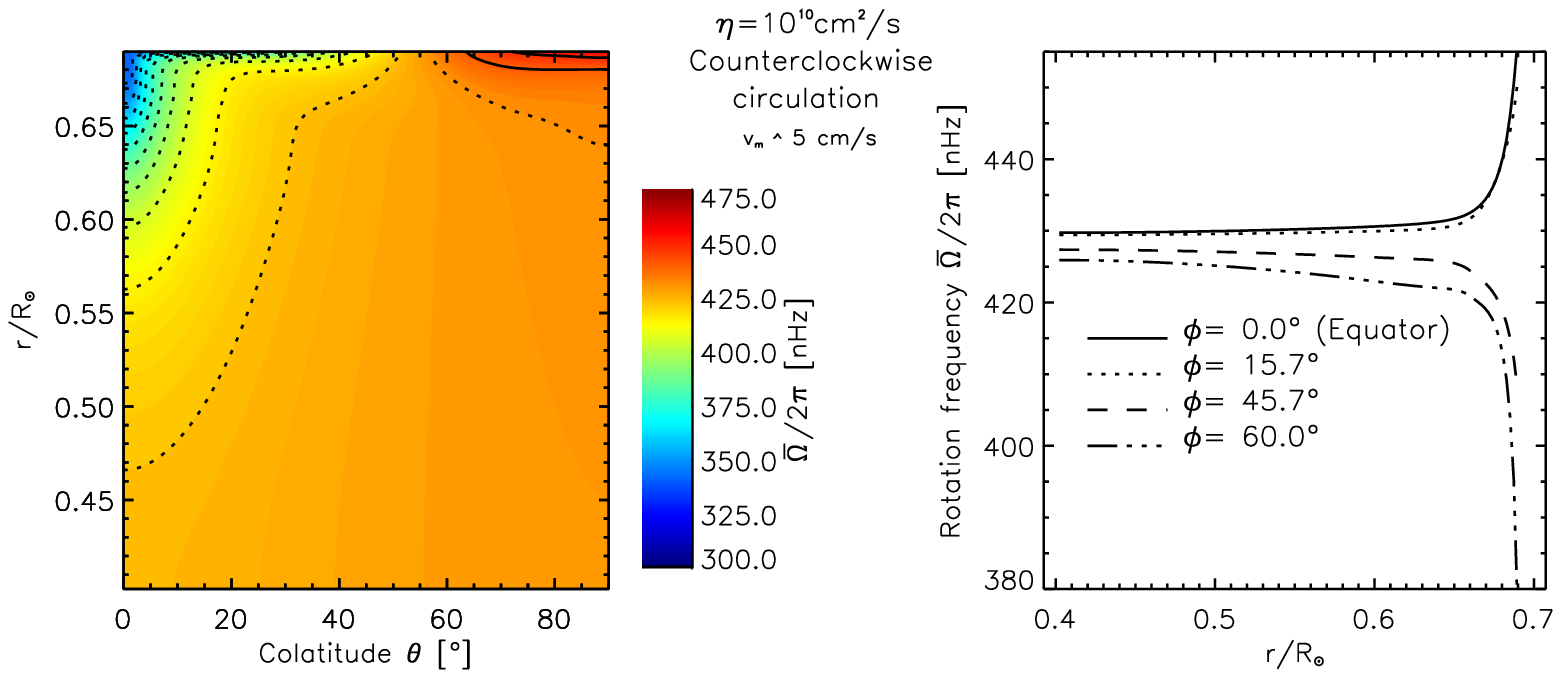}
   \includegraphics[width=0.49\linewidth]{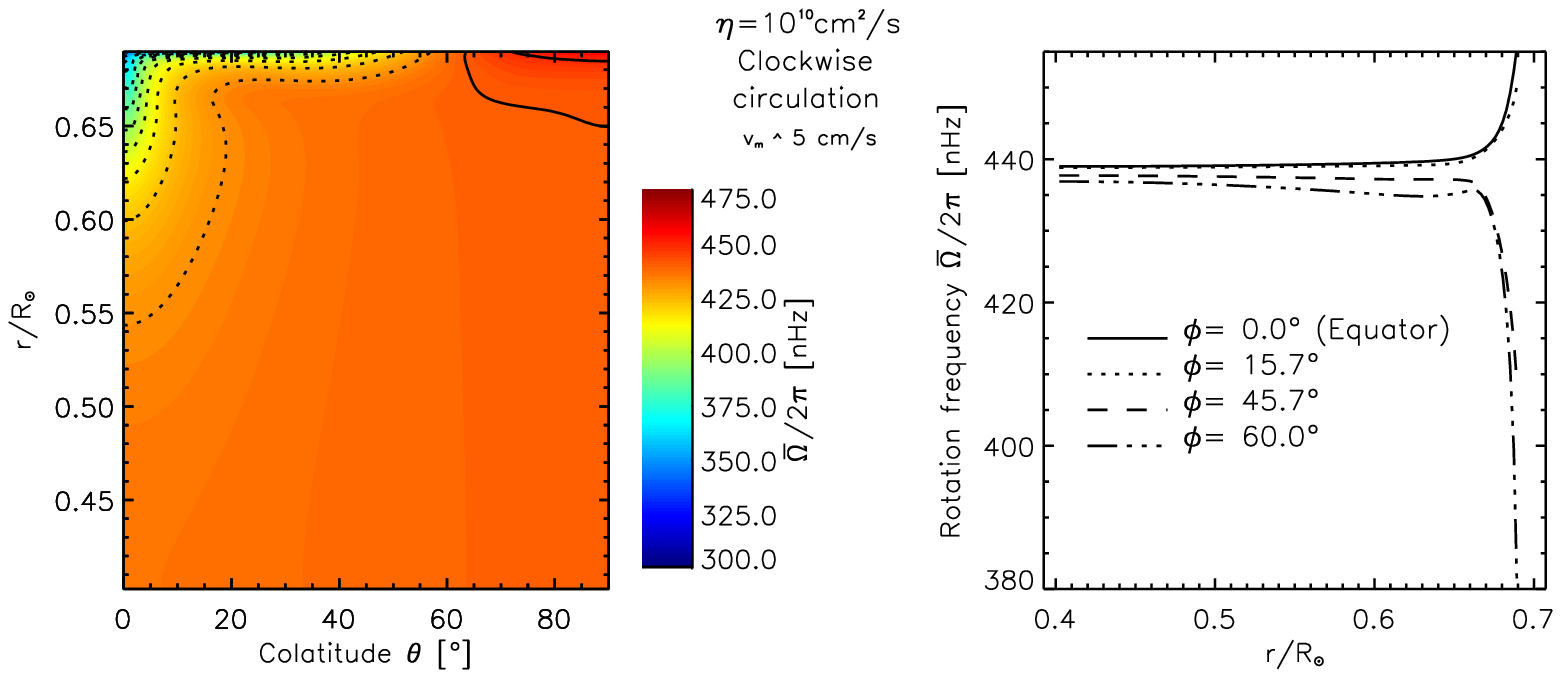}
      \caption{Same as in Fig.\,\ref{fig:1_atlag_col}, but including the 
               meridional circulation. In {\it left-hand panel} the circulation is counterclockwise, 
	       while in {\it right-hand panel} it is clockwise.
              }
         \label{fig:2_atlag_col}
   \end{figure*}

\section{Results}

As stated above, we examine the influence of a migrating field on the 
radial spreading of the differential rotation into the radiative interior. 
Prescribing a migrating magnetic field at the top we define the 
form of the vector potential from the observed radial magnetic field. 
Besides the effect of the field strength, we study the consequences of the 
varying the diffusivity and the magnetic Prandtl number with radius, and the 
influence of the meridional circulation.

\subsection{Solution Without Meridional Circulation}

In the first calculations, only the field strength varies from case to case. 
The values of the viscosity and magnetic diffusivity chosen for these 
simulations are identical, $\eta=\nu=10^{10}
\mbox{cm}^2/\mbox{s}$, i.e. the magnetic Prandtl number is one $\Prm=\nu/\eta=1$ and 
the meridional circulation is neglected. Several calculations were run with 
different field strengths and we chose the most suitable case for us, 
where the sufficiently strong 
magnetic field is able to interfere the spreading of the 
differential rotation into the radiative interior and to reproduce the observed 
thickness of the tachocline. This case is shown in Fig.\,\ref{fig:1_atlag_col} 
after relaxation 
               \footnote{Color figures and computer animations illustrating the time development of some 
	       of our solutions can be downloaded from the following web site: 
	       {\tt http://astro.elte.hu/kutat/sol/fast2/fast2e.html}}. 

In the left-hand panel 
we plotted the contours of the time-average of the rotation rate, which 
is defined as
\begin{eqnarray}
\overline{\Omega}(r,\theta)&=&\overline{\omega}(r,\theta)+\Omega_0 \\
\overline{\omega}(r,\theta) &=&\frac {1}{\Pcyc}\int_t^{t+\Pcyc}{\omega(r,\theta,t)}\,\mathrm{d}t,
\end{eqnarray}
where $\Pcyc=22$yr is the period of the cycle.

In the right-hand panel we plotted the rotation rate at different latitudes as 
a function of radius. 
We note that in \citetalias{FD+P:AA02} we plotted the differential rotation amplitude 
$\Delta\omega$ in right-hand panels, where $\omega$ was weighted with latitude 
and $\Delta\omega$ was normalized (see Eq.~(23) in \citetalias{FD+P:AA02}). 
However, in this paper we retained
the values of the frequency, so in this case the results can be compared easier 
with helioseismic observations and even the more insignificant 
variations of the rotation rate are more conspicouos.
Based on the helioseismic observations we expect that the 
differential rotation changes to the rigid rotation in a thin transition region. 
It is well visible that the prescribed differential rotation at the base of the 
convection zone changes to the near-uniform rotation in a thin layer. 

We also study in detail the dynamics of the tachocline and compare to the observations. 
Accordingly, the thickness of the tachocline and the change in the rotation rate across 
the tachocline are defined as in \citet{Basu:MNRAS97}. 
Thus, the jump in the rotation rate across the tachocline $\delta\Omega$ is the 
difference between the rotation rate at the base of the convection zone and the 
rotation rate in the interior. The thickness of the tachocline $w$ is defined as
the rotation rate increases from the factor $1/(1+e)$ of its maximum value to 
the factor $1-1/(1+e)$ of its maximum value in the range $r=\rt-w$ 
to $r=\rt+w$. 

   \begin{figure}
   \centering
   \includegraphics[width=0.7\linewidth]{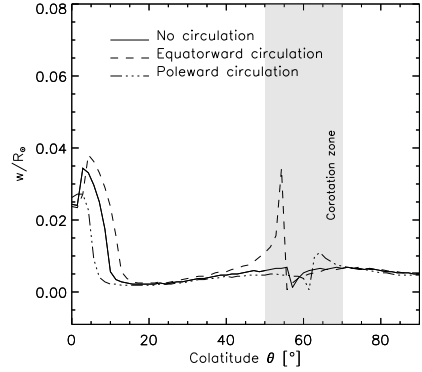}
      \caption{Latitudinal variation of the tachocline thickness. The solid line 
               represents the case shown in Fig.\,\ref{fig:1_atlag_col} and 
	       the dashed and dashed-dot lines respectively denote the thickness 
	       including the meridional circulation.
	      }
         \label{fig:w_th_3sum}
   \end{figure}

The fast tachocline shows a quite marked latitude-dependence 
in our results, as illustrated in Figs.\,\ref{fig:1_w_th}-\ref{fig:1_jump_th}.
In both figures we also plot the results of the helioseismic measurements 
from MDI and GONG \citep{Basu+Antia:MNRAS01}. It should be noted that, according to the 
helioseismic results, the latitudinal variation in position of tachocline is 
significant, while the variation in width is not clear, but the results tend 
to suggest that the thickness increases with latitude. Since  
the top of the computational domain is fixed at $0.69 R_\odot$ in our model, 
therefore the position of the tachocline cannot be directly compared with 
the observations.  
Notice that the thickness of the tachocline is close to zero at around the colatitude of 
$60^\circ$, because at this colatitude the rotation rate is equal to  
the rigid rotation velocity. This region is the so-called corotation zone (Fig.\,\ref{fig:1_w_th}). 
In addition to latitudinal variations, we also investigate the temporal variations 
in the properties of the tachocline, and we find that the thickness of the tachocline 
depends on cycle phase (Fig.\,\ref{fig:1_w_time}). 
This variation is not confirmed by the observations, but the helioseismic data set is 
relatively short.
The development of the toroidal magnetic field 
and the radial magnetic field shown in Fig.\,\ref{fig:1_tor_rad_time} correspond to 
the observed evolutionary pattern of the magnetic field, i.e. the butterfly diagram. 

\subsection{Solution With Weak Meridional Circulation}

In the second case, we study the influence of the meridional circulation. 
As we mentioned, the strong subadiabatic stratification sets an upper limit to 
the amplitude of the meridional circulation ($v_m \sim 10 \mbox{ cm s}^{-1}$). 

We use the flow patterns shown in \citetalias[][Figs.~9 and 10]{FD+P:AA02}, where the speed of the 
horizontal flow is weak.
In these cases, we considered the 
effect of the meridional flow on the magnetic field 
(M$_2$, M$_3$ in Eqs.\,(\ref{eq:main2}-\ref{eq:main3})). 
The results with the meridional circulation included can be seen in 
Fig.\,\ref{fig:2_atlag_col}. The poloidal field amplitude used 
here was the same as in the previous case.
Notice that the effect of the weak meridional flow of $\sim 5 \mbox{ cm s}^{-1}$ 
on the magnetic field is not dramatic.

In Fig.\,\ref{fig:2_atlag_col} it can be seen 
that the rigid rotation rate is shifted compared to the first case in 
accordance with the direction of flow. An equatorward flow in the tachocline region 
increases the near-uniform rotation rate, while a poleward flow reduces it.
We remark that the lower boundary condition for $\omega$ is different than in 
\cite{FD+P:SolPhys01,FD+P:AA02} where $\omega=0$ at the bottom of our box. 
In these cases $\partial_r\omega=0$ at the bottom 
of the computational domain, so the effect of the meridional circulation 
on the rotation is perceptible in the radiative interior. It is apparent that the 
deep meridional flow changes the rigid rotation rate. 
If we raised the amplitude of the meridional circulation at the upper part of the 
radiative interior ($v_m \sim 2 \mbox{ m s}^{-1}$), but we reduced the penetration 
of the meridional circulation -- the meridional motion is in the overshoot region --, 
the shift is not dramatic. This result could play an important role in 
the model for the solar dynamo models wherein a meridional flow penetrates 
into the stable layers below the tachocline \citep{Nandy+Choudhuri:Sci02}. 

In Fig.\,\ref{fig:w_th_3sum}\,the differences between the three results are apparent. 
Note that the rotation rate in the interior is changed in these cases and we used it to define 
the thickness of the tachocline.
It is well visible that the effect of the weak circulation is not dramatic and 
chiefly limited to the pole. In this paper the toroidal field strength is higher as in 
the earlier paper and on the other hand the upper boundary condition 
on the poloidal field is controlling the influence of the meridional motion. 
In case of equatorward circulation it can be seen that the tachocline thickness at the pole  
extended and in the corotation zone the peak is apparent. Both are due to the transport by 
the equatorward motion. If we assume that in the tachocline region the circulation is 
toward the pole then the thickness at the high latitudes is reducible. It must be noted 
of course that the solar dynamo theories favour the counterclockwise circulation. 
On the other hand in our model we have not emphasized the role of the circulation, but 
the role of the Maxwell stresses which are capable to limit the shear layer to its 
observed thickness. We studied it for the sake of completeness.

\subsection{Varying the diffusivity with radius and $\Prm=1.0$}

   \begin{figure*}[!ht]
   \centering
   \includegraphics[width=0.49\linewidth]{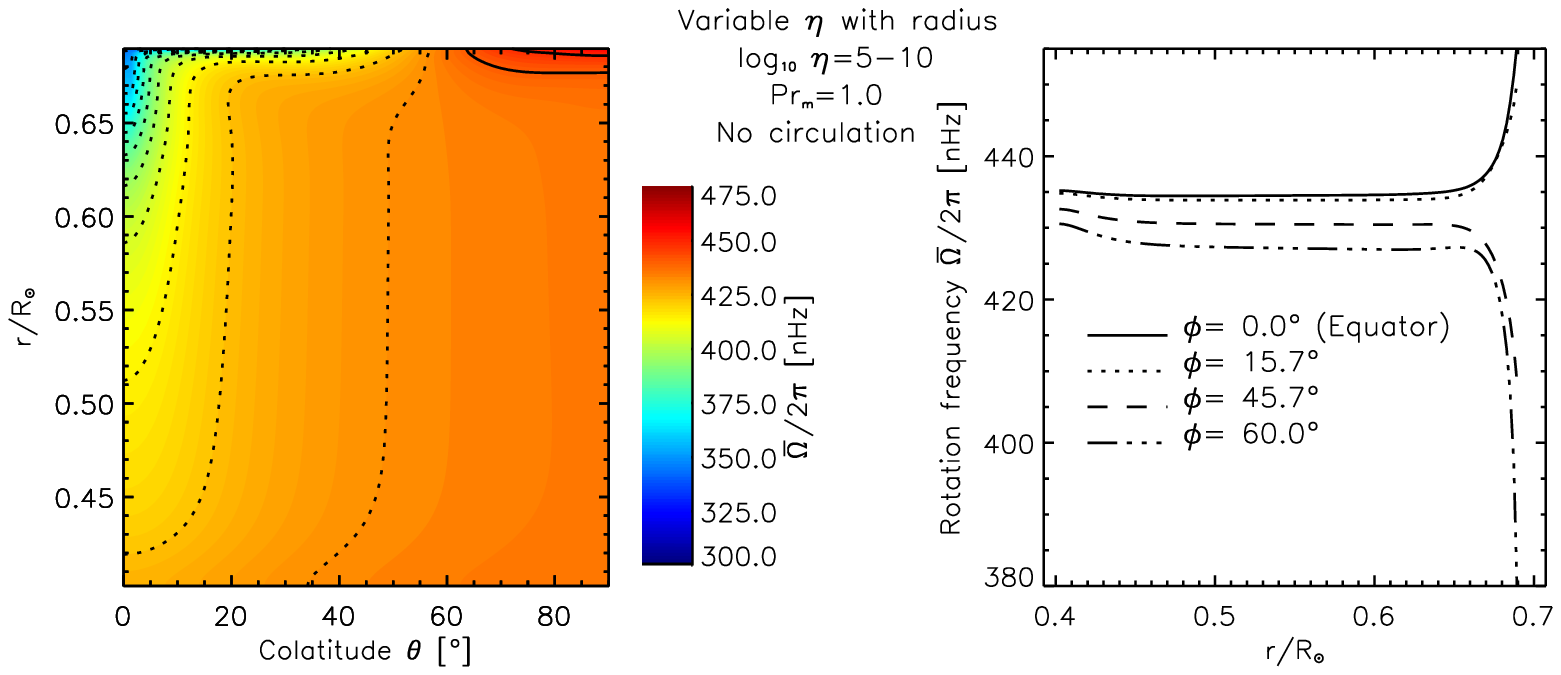}
   \includegraphics[width=0.49\linewidth]{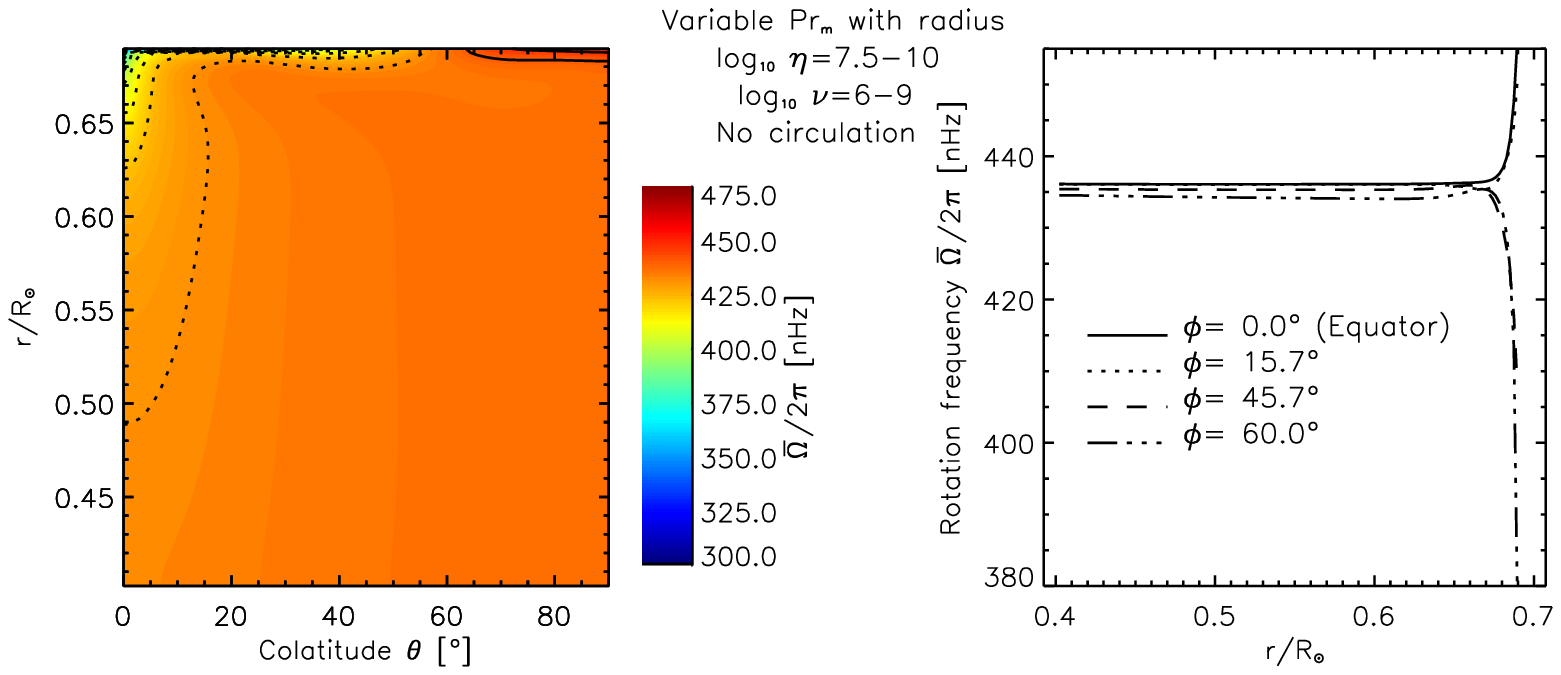}
      \caption{Same as in Fig.\,\ref{fig:1_atlag_col}. {\it Left-hand panel:} the diffusivity is 
               varied with radius. 
	       {\it Right-hand panel:} the magnetic Prandtl number is 
               varied with depth.
              }
         \label{fig:4_atlag_col}
   \end{figure*}

In this section the diffusivity is allowed to vary with radius. In 
\citetalias{FD+P:AA02} we only investigated cases where the 
diffusivities are constant throughout the computational domain. 
The question may arise that the high viscosity in the deep 
radiative zone could be a disfigurement to our results? In order to 
answer, we change the magnetic diffusivity and the viscosity with radius as 
\begin{eqnarray}
\label{eq:eta_r}
\eta \,(\mbox{or } \nu)=\eta_0\, (\mbox{or } \nu_0)\left[1+\exp\left(\Gamma\frac{r_v-r}{\rbcz-\rin} \right) \right]^{-1},
\end{eqnarray}
where $\eta_0$, $\nu_0$ is the maximum value of the diffusivity and viscosity, 
$\Gamma$ and $r_v$ are geometric parameters.  
Important notice that the used numerical technique sets a limit to this  
variation with radius. Namely, the lowest value of viscosity or 
magnetic difusivity in the domain determines the diffusive timescale as against 
the time step which is determined by the highest value of these. Hence the 
very low value of viscosity gives rise to the extremly long integration times.

Thus, in this case we set these parameters 
as follows: $\eta_0\, (\mbox{or } \nu_0)=10^{10}$, $\Gamma=16$ and $r_v=4.2\times10^{10}$, 
and the magnetic Prandtl number is one $\Prm=\nu/\eta=1$. 
The magnetic diffusivity is varied with depth between approx. $10^5$ and $10^{10}$.
This variation of the diffusivity, 
while quantitatively much milder than expected physically, provides an opportunity for us 
to consider the effects of a variable diffusivity without the necessity of impossibly 
long integration times.

In left-hand panel of Fig.\,\ref{fig:4_atlag_col} it is clearly visible that the 
influence of the downwards decreasing diffusivity on the solution is small. 
However it is noticeable that in the polar region the differential rotation 
penetrates into deeper layers of the radiative interior. On the other hand 
the amplitude of the temporal variation of the thickness is smaller than in 
the case of Fig.\,\ref{fig:1_w_time}.

\subsection{Varying the magnetic Prandtl number with depth}

Under realistic conditions, the turbulent magnetic diffusivity in the tachocline 
should be reduced to the microscopic magnetic diffusivity in the radiative core and 
the turbulent viscosity to molecular and radiative viscosity. However the magnetic 
diffusivity is greater than the viscosity and it is less reduced than the viscosity 
\citep[see esp. Fig.\,1 in][]{Rudiger+Kichatinov:AN97}, so 
the magnetic Prandtl number is varied with depth. 

As argued earlier, the low diffusivities often make the numerical simulations 
problematic, therefore we tried to give a qualitative test instead of a 
quantitative analysis. Accordingly, 
we used the above formula (\ref{eq:eta_r}) to define the variation of 
$\eta$ and $\nu$ with radius and 
the required parameters are following: $\eta_0=1\times10^{10}$, $\Gamma=8$, 
$r_v=4.2\times10^{10}$ for $\eta$  and $\nu_0=10^{9}$, $\Gamma=10$, 
$r_v=4.2\times10^{10}$ for $\nu$, so the magnetic Prandtl number
is varied between 
0.024 and 0.1.

\begin{figure}
   \centering
   \includegraphics[width=0.7\linewidth]{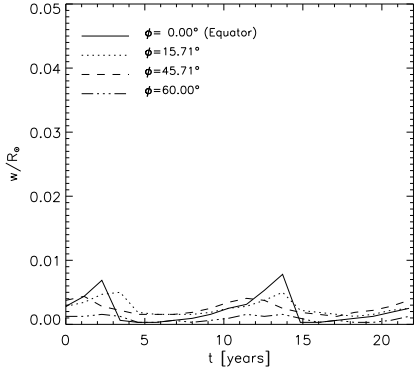}
      \caption{The thickness of the tachocline at different latitudes as a 
               function of time
	       for the case in right-hand panel of Fig.\,\ref{fig:4_atlag_col} 
	       (varying the magnetic Prandtl number with depth).
	       }
         \label{fig:5_w_time}
   \end{figure}

Our result is illustrated in right-hand panel of Fig.\,\ref{fig:4_atlag_col}. In this case 
the poloidal field strength is $B_p\sim 1500$ G and the maximum value of the 
toroidal field is $B\sim 6000$ G. It is well visible that 
the penetration of the differential rotation into the radiative interior is strongly reduced 
in the polar region and the temporal variation of the thickness 
is also decreased (see Fig\,\ref{fig:5_w_time}). 

\section{Discussion and Conclusion}

The models discussed in this paper show that the dynamo field dominates over 
the dynamics of the fast solar tachocline provided that the turbulent diffusivity 
in the tachocline region exceeds 
$\eta\ga10^9\,$cm$^2/$s.

In our first work \citep{FD+P:SolPhys01} we presented an analytical estimate for 
the strength of an oscillatory poloidal field, which is able to confine 
the tachocline to its observed thickness for $\eta\sim 10^{10}\,$cm$^2/$s. 
Based on the estimate we showed a simple numerical calculation. 
A crucial assumption in this calculation is that the oscillatory poloidal 
magnetic field does penetrate below the convective zone to depths comparable to
the tachocline thickness and the penetration is prescribed a priori. 

In our forthcoming model (\citetalias{FD+P:AA02}), instead of the 
a priori prescribed oscillatory poloidal 
magnetic field we included the evolution equation for the poloidal field, but 
the poloidal field was a simple oscillating dipole, which was not very realistic. 
In that work we also explored the parameter space a bit more widely by varying 
the magnetic Prandtl number and the diffusivity, but we did not investigate the 
effects of the variations of these parameters with depth. We found that the 
confining field strength increased with increasing viscosity and magnetic Prandtl number. 

In the present work we focus on three shortfalls of the earlier models. First, instead of 
the simple oscillating dipole poloidal field we study the more general magnetic 
field structures reminiscent of the butterfly diagram. On the other hand, a more 
realistic model should have a magnetic diffusivity decreasing significantly inside 
the radiative interior. 

For the more general magnetic field structures, we use the results of 
\citet{Stenflo:ASS88,Stenflo:SSM94}, where 
the evolutionary pattern of the axisymmetric magnetic field can be written 
as a superposition of the sinusoidal, $22$ yr variations for the $7$ odd 
modes spherical harmonic components. Based on our previous works we studied 
the field strength and we found that a sufficiently 
strong migrating magnetic field is able to reproduce the observed thickness 
of the tachocline. This agrees with the results in \citet{FD+P:SolPhys01,FD+P:AA02}, 
apart from the fact that the time-latitude diagram for the toroidal field 
(Fig.\,\ref{fig:1_tor_rad_time}) is now similar to the butterfly diagram. 
The lower boundary condition for $\omega$ is different than in 
\cite{FD+P:SolPhys01,FD+P:AA02}, giving us a possibility to 
study the effect of the dynamics of the upper layers on the inside layers. 
In Fig.\,\ref{fig:1_atlag_col} it can be seen that because of the high diffusivity
there is a little deviation in the rotation rate compared to the fixed internal 
rotation rate $\Omega_0$ at different latitudes. Furthermore an equatorward 
meridional circulation in the tachocline region 
increases the near-uniform rotation rate, while a poleward flow reduces it. 
Note that this phenomenon is sensitive to the penetration depth of the flow into 
the upper radiative interior. We found that the internal rotation rate is not 
shifted extremly in spite of the increasing amplitude of circulation if the 
circulation is dominated in the upper part of the tachocline. 
The downwards decreasing diffusivity or rather the magnetic Prandtl number does not 
appreciably affect the solution. Evidently, this model is simplified treatment of the 
turbulence in the tachocline, nevertheless it gave us an opportunity to consider 
these effects. The varying magnetic diffusivity and Prandtl number with depth 
give rise to a change of the temporal variation of the tachocline thickness. 
It is shown in Fig.\,\ref{fig:5_w_time} 
that the downwards decreasing magnetic Prandtl number tends to reduce the 
spatiotemporal variation. 

\begin{figure}
   \centering
   \includegraphics[width=0.7\linewidth]{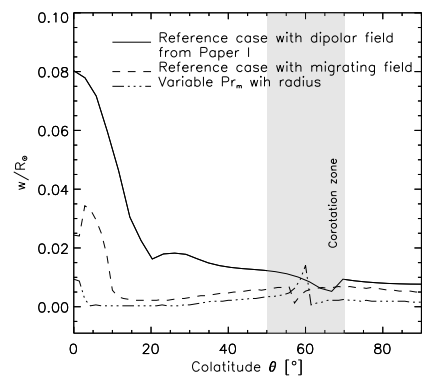}
      \caption{Latitudinal variation of the tachocline thickness. The solid line 
               represents the case shown in Fig.6 of \citetalias{FD+P:AA02} and 
	       the dashed and dashed-dot lines respectively denote the thickness 
	       including the meridional circulation.
	       }
         \label{fig:w_th_sum_paper1}
   \end{figure}

The main results of present and earlier models is that the dynamo field 
in the turbulent tachocline by Maxwell stresses is capable to limit the 
shear layer to its observed thickness. The next step of this analysis consists in 
comparing the numerical results in the present work and the model in 
\citetalias{FD+P:AA02} with the helioseismic results. In \citetalias{FD+P:AA02} 
the thickness of the tachocline showed the ''pit'' at the pole, which is not confirmed 
by the available helioseismic information, which may be in consequence of 
the latitudinal limit of observations. 
In \citetalias{FD+P:AA02} the poloidal field was a simple oscillating dipole, 
and the reality of this pit needs to be verified with other poloidal field geometries and 
other parameters. In this paper we found that in case of the more general magnetic field 
structure the polar pit is reduced to its half and in case of varying the magnetic 
Prandtl number with depth the pit is almost disappeared (Fig.\,\ref{fig:w_th_sum_paper1}). 
If we use the more 
general field geometry reminiscent of the butterfly diagram, the penetration depth of 
the differential rotation into the radiative interior in the polar region is decreased 
owing to the fact that at higher latitudes the stonger magnetic fields stay up.
In case of the decreasing magnetic Prandtl number the influence of the Maxwell stresses 
are in the ascendant compared with the viscous stresses, 
so the permeation of the differential rotation below the convective 
zone at the pole reduces further. 
On the other hand regarding the dynamical aspect, we studied the temporal variations of 
tachocline. Due to the relatively short helioseismic data set, the temporal variations 
associated with the solar cycle to appear in the tachocline is not clear. In both models,  
the vector potential of the magnetic field is prescribed with a period of 22 years at the 
top of the computational domain, so we found that the thickness of the tachocline depends 
on cycle phase. However the amplitude of the temporal variations can be reduced if the impact of 
Maxwell stresses is enhanced compared with the viscous stresses -- for example the 
magnetic Prandtl number is less than one (cf.~Fig.\,\ref{fig:5_w_time}).

As stated earlier, an important shortfall of these models 
is  obviously the extremely simplified treatment of the turbulent transport. 
If the turbulence arises from the instabilities in the tachocline,
then the diffusivities should clearly be some, possibly very complex,
functionals of the velocity field and of the magnetic field. The next step to
model this would be to use a Smagorinsky-type formula or $k-\epsilon$ model 
for the diffusivities.

Other important points what we should consider in the future are the boundary conditions 
on the magnetic field and the 
complete dynamo-generated poloidal field, because the conditions in the present treatment poorly 
represent the 
dynamical interaction between the flow and the magnetic field in the tachocline and 
the convection 
zone \citep{Petrovay:SOLSPA}. Many theoretical models suggest that the toroidal magnetic 
field is up to 
$10^5$ G in the tachocline, required to produce sunspots in low latitudes \citep{Petrovay+Moreno:1997}. 
In our model the peak 
toroidal field is above about $10^4$ G, but this may change if different boundary 
conditions 
are used or the dynamo-generated field is included. \citet{Dikpati+Gilman:ApJ01,Gilman+Dikpati:ApJ02} 
suggest that the dynamo-generated toroidal field 
should play an important role in the global instability of tachocline differential 
rotation. 
In future work we plan to develop our model along the lines described above.

\acknowledgements 
I would like to thank J.~O. Stenflo for providing the parameters of the evolutionary pattern of the 
axisymmetric radial magnetic field, K. Petrovay for helpful discussions on the manuscript and 
my colleagues (T. Borkovits and V. K\"onyves) for their help. 
This work was supported in part by the OTKA under grant no.~T043741.


%


\begin{thebibliography}{}

\bibitem[Basu(1997)]{Basu:MNRAS97}
Basu, S. 1997, \mnras, 288, 572

\bibitem[Basu \& Antia(2001)]{Basu+Antia:MNRAS01}
Basu, S., Antia, H.~M. 2001, \mnras, 324, 498

\bibitem[Brun et~al.(1999)]{Brun+:ApJ99}
Brun, A.~S., Turck-Chi\`eze, S., Zahn, J.-P. 1999, \apj, 525, 1032

\bibitem[Brun et~al.(2002)]{Brun+:AA02}
Brun, A.~S., Antia, H.~M., Chitre, S.~M., Zahn, J.-P. 2002, \aap, 391, 725

\bibitem[Charbonneau et~al.(1999)]{Charbonneau+:ApJ99}
Charbonneau, P., Christensen-Dalsgaard, J., Henning, R., et~al. 1999,
\apj, 527, 445

\bibitem[Corbard et~al.(1999)]{Corbard+:AA99}
Corbard, T., Blanc-Féraud, L., Berthomieu, G., et~al. 1999, \aap,
344, 696

\bibitem[Corbard et~al.(2001)]{Corbard+:SOGO}
Corbard, T., Jim\'enez-Reyes, S.~J., Tomczyk, S., et~al. 2001, in Helio- and Asteroseismology at the Dawn of the
Millennium (ESA Publ. SP-464), p.~265

\bibitem[Dikpati \& Gilman(2001)]{Dikpati+Gilman:ApJ01}
Dikpati, M, Gilman, P.~A. 2001, \apj, 559, 428

\bibitem[Elliott \& Gough(1999)]{Elliott+Gough:ApJ99}
Elliott J.~R., Gough D.~O. 1999, \apj, 516, 475

\bibitem[Forg\'acs-Dajka \& Petrovay(2001)]{FD+P:SolPhys01}
Forg\'acs-Dajka, E., Petrovay, K. 2001, \solphys, 203, 195

\bibitem[Forg\'acs-Dajka \& Petrovay(2002)]{FD+P:AA02}
Forg\'acs-Dajka, E., Petrovay, K. 2002, \aap, 389, 629 (Paper I)

\bibitem[Garaud(2001)]{Garaud:MNRAS01}
Garaud, P. 2001, \mnras, 329, 1

\bibitem[Gilman(2000)]{Gilman:SolPh00}
Gilman, P.~A. 2000, \solphys, 192, 27

\bibitem[Gilman \& Dikpati(2002)]{Gilman+Dikpati:ApJ02}
Gilman, P.~A., Dikpati, M. 2002, \apj, 576, 1031

\bibitem[Gough \& McIntyre(1998)]{Gough+McIntyre:Nat98}
Gough, D.~O., McIntyre, M.~E. 1998, Nature, 394, 755

\bibitem[Howe et~al.(2000)]{Howe+:Sci00}
Howe, R., Christensen-Dalsgaard, J., Hill, F., et~al. 2000, \Sci, 287, 2456

\bibitem[{Kitchatinov \& R{\"u}diger(1995)}]{Kitchatinov+Rudiger:AA95}
Kitchatinov, L.~L., R{\"u}diger, G. \&  1995, \aap, 299, 446

\bibitem[Kosovichev(1996)]{Kosovichev:ApJ96}
Kosovichev, A.G. 1996, \apj, 469, L61

\bibitem[{K\"uker \& Stix(2001)}]{Kuker+Stix:AA01}
K\"uker, M., \& Stix, M. 2001, \aap, 366, 668

\bibitem[MacGregor \& Charbonneau(1999)]{McGregor+Charbonneau:ApJ99}
MacGregor, K.~B., Charbonneau, P. 1999, \apj, 519, 911

\bibitem[Makarov \& Sivaraman(1989)]{Makarov+Sivaraman:SolPhys89}
Makarov, V.~I., Sivaraman, K.~R. 1989, \solphys, 123, 367

\bibitem[Miesch et~al.(2000)]{Miesch+:ApJ00}
Miesch, M.~S., Elliott, J.~R., Toomre, J., et~al. 2000, \apj, 532, 593

\bibitem[Nandy \& Choudhuri(2002)]{Nandy+Choudhuri:Sci02}
Nandy, D., Choudhuri, A.~R. 2002, \Sci, 296, 1671

\bibitem[Petrovay \& Moreno-Insertis(1997)]{Petrovay+Moreno:1997}
Petrovay, K. and Moreno-Insertis, F. 1997, \apj, 485, 398

\bibitem[Petrovay \& Szak\'aly(1999)]{Petrovay+Szakaly:1999}
Petrovay, K. and Szak\'aly, G. 1999, \solphys, 185, 1

\bibitem[Petrovay(2000)]{Petrovay:SOLSPA}
Petrovay, K. 2000, in The Solar Cycle and Terrestrial Climate, ESA Publ. SP-463,
p.~3; also astro-ph/0010096

\bibitem[R{\"u}diger \& Kitchatinov(1997)]{Rudiger+Kichatinov:AN97}
R{\"u}diger, G., Kitchatinov, L.~L. 1997, \an, 318, 273

\bibitem[Schou et~al.(1998)]{Schou+:ApJ98}
Schou, J., Antia, H.~M., Basu, S., et~al. 1998, \apj, 505, 390

\bibitem[Spiegel \& Zahn(1992)]{Spiegel+Zahn:AA92}
Spiegel, E.~A., Zahn J.-P. 1992, \aap 265, 106

\bibitem[Stenflo(1988)]{Stenflo:ASS88}
Stenflo, J.~O. 1988, \ASS 144, 321

\bibitem[Stenflo(1994)]{Stenflo:SSM94}
Stenflo, J.~O. 1994, in R.~J. Rutten and C.~J. Schrijver (eds.), Solar Surface Magnetism, p.~365

\end{thebibliography}
\end{document}